\documentclass[prd,twocolumn,preprintnumbers,aps]{revtex4}%
\usepackage{epsfig}
\usepackage{amsmath}
\usepackage{amsfonts}
\usepackage{amssymb}
\usepackage{graphicx}%
\providecommand{\U}[1]{\protect\rule{.1in}{.1in}}

\def\be{\begin{equation}}
\def\en{\end{equation}}
\def\bq{\begin{eqnarray}}
\def\eq{\end{eqnarray}}
\begin{document}
\preprint{ }
\title{Static Universe model existing due to the Matter-Dark Energy coupling}
\author{A. Cabo Bizet$^{1}$ and A. Cabo Montes de Oca $^{2,3}$}
\affiliation{$^{1}$ \textit{ Facultad de Fisica de la Universidad de la Habana}}
\affiliation{$^{2}$ \textit{Grupo de F\'{\i}sica Te\'{o}rica, Instituto de Cibernetica,
Matem\'{a}tica y F\'{\i}sica, Calle E No 309, Vedado, La Habana, Cuba}}
\affiliation{$^{3}$ \textit{Abdus Salam International Centre for Theoretical Physics, 
Strada Costiera 111, Miramare, Trieste, Italy }}

\begin{abstract}
\noindent The work investigate a static, isotropic and almost homogeneous
Universe containing a real scalar field modeling the Dark-Energy
(quintaessence) interacting with pressureless matter. It is argued that the
interaction between matter and the Dark Energy, is essential for the very
existence of the considered solution. Assuming the possibility that
Dark-Energy can be furnished by the Dilaton (a scalar field reflecting  the
condensation of string states with zero angular momentum) we fix the value of
scalar field at the origin to the Planck scale. \  It became  possible to
\ fix the ratio of the amount of Dark Energy to matter energy,  in  the
currently estimated value $\frac{0.7}{0.3}$ and also the  observed  magnitude
of the Hubble constant. The small valueof the mass for the scalar field chosen
for fixing the above ratio and Hubble effect strength, results to be of the
order of \ $10^{-29}cm^{-1}$, a small value which seems to be compatible with
the zero mass of the Dilaton in the lowest approximations.

\noindent

\end{abstract}
\maketitle

\section{Introduction}

The assumption about the isotropic and homogeneous nature of our
Universe, that is the Cosmological Principle, is central to modern
Cosmology \cite{Weinberg}. However, recent experimental observations
suggest the possibility for the break down of the validity of the
principle at large scales \cite{Alnes}. En el present work a static
solution of the KG equation in interaction with matter is
investigated which shows a large region of homogeneity close to a
central symmetry point, but not at large distances. The existence of
this static solution essentially rests on the presence of an
interaction of the scalar field (modeling the quintessence) with the
pressureless matter. The solution discussed here is a generalization
of one formerly investigated in Ref. [\onlinecite{papa2,papa1}] in
the absence of matter. The special characteristics of the scalar
field led to the proposal made in Ref. [\onlinecite{papa1}] of
considering it as representing the Dilaton of the string theory
\cite{witten}. This idea came from the fact that when you fix the
value of the scalar field (which have dimension of mass) at the
central symmetry point at the Planck scale, by also requiring an
amount of Hubble effect similar to the experimental one, the radius
of existence of the solution gets a value near to $R=10^{28}cm$. \
However, even more curious is that the values of KG mass of the
fields obtained by fixing the above parameters, results to be of the
order of $1/R$. That is, a  very small value which seems compatible
with a very tiny mass acquired by the Dilaton due to boundary
conditions or non perturbative effects, \ which could deviate it
from its known massless character in the first approximation.

Various static models of the Universe have been considered. Among
them are the ones of Einstein, Le Maitre and de'Sitter,
respectively. Originally, Einstein \cite{Weinberg} examined a
Universe filled of uniformly distributed matter but obtained a
non-static metric. This result motivated him to introduce in his
equations the Cosmological Constant term $\lambda$, with the
objective of allowing  the obtaining of a static solution. However,
this model contradicted the observations at that time and was
disregarded.

The de'Sitter \ model considers the Universe as empty and is defined
by a Cosmological Constant $\lambda\equiv\frac{3}{r_{s}^{2}}$. The
metric has the form:
\begin{align}
ds^{2}  & =(1-\frac{r^{2}}{r_{s}^{2}}){cdt}^{2}-\frac{1}{(1-\frac{r^{2}}%
{r_{s}^{2}})}dr^{2}-\nonumber\\
& r^{2}(sin^{2}\theta d\varphi^{2}+d\theta^{2}),\nonumber
\end{align}
which describes an homogeneous space of volume
\[
V=2\pi^{2}r_{s}^{3}.
\]
\ The high symmetry of this space-time is evidenced by the property
that any point of it can be transformed in the origin of coordinates
by a bijective transformation \cite{Susskind}. However, the fact
that in this model  the Universe is empty,  is unreal.\ \ It can be
underlined that experimentally and theoretically the DE has been
associated to the presence of \ a cosmological constant [\cite{Perl}
]. However, in this modern view the Cosmological Constant coexists
with matter.\

In connection with the above fact, it has been observed that a
centrally symmetric static scalar field \ satisfying the
Einstein-Klein-Gordon equations (EKG) \ curves the space time \ in a
form resembling  the one in the de'Sitter space in a large
neighborhood of the origin of coordinates \cite{papa2}.  The fact
that the scalar field is more weakly varying along the radial domain
when its  value at the origin is lower is  an interesting property
to notice. The associated densities of energy and pressure are
positive and \ negative respectively and weakly varying,
approximating the  presence of a positive Cosmological Constant.
These properties suggested the idea advanced in \cite{papa1} of
considering the Dark Energy (DE) as described by a scalar field in
this approximately homogeneous field configuration studied in
\cite{papa2}. \ This assumption will determine the abandoning of the
Cosmological Principle in favour of what could be imagined as a kind
of "Matryoshka" \ model of the Universe. In this conception,
proposed in \cite{papa2,papa1}, we could be living inside of a
particular configuration in which the Dilaton field has a definite
value resulted from the collapse of \ string matter in fermionic
states. The presence of the DE is then assumed to be associated to
the Dilaton field which could be radiated by the string matter in
fermionic states under the extreme conditions of the collapse. \ \
The effective realization of this picture in  Nature, could lead to
the possibility that the astrophysical black-holes \ (by example the
ones which are expected to exist near the centres of the Galaxies)
could be no other things that small Universes in which the Dilaton
field \ gets a different value to the external one associated to the
DE. This change could be \ produced again by the collapse of \
fermion matter in falling to the hole,  upon the possible radiation
of zero angular momentum modes, that is of the Dilaton. \ \ We
encounter this picture as interesting and think its exploration is
worth considering.

Therefore, in this work we address the finding of a static solution
\ related with the one discussed in \cite{papa2,papa1}, \ but also
including pressureless matter,  in order to approach the discussion
of the physical properties of these configurations in a situation
more closely related with the physical conditions in the Universe.

An important aspect which  emerged in the first examination of the
problem, is that the coexistence of the scalar field as described by
the EKG equations including also the dust \ energy momentum tensor
does not allow the existence
of static solutions, at least in centrally symmetric configurations \cite{thoft}%
. However, in this work we will \ argue that such a solution exists
if the interaction between the matter and the Dark Energy (or
Dilaton condensate) is \ included. \ The introduction of the
coupling does not damage the almost homogenous character of the
solution in a relative large region around the origin of the central
symmetry, being far away form the limits of the Universe. An
interesting outcome is that the distributions of \ matter and of
Dark Energy both show a very close  behavior. That is, the scalar
field (Dark Energy) is able to \ sustain an amount of matter being
almost proportional between them.

The hypothesis of a pressureless matter (Dust Filled Universe)
\cite{thoft} is employed here. It constitutes a reasonable
assumption in the framework of modern Cosmology \cite{Weinberg},
reflecting the idea that the astrophysical systems are formed by
large collections of galaxies \ and these again are grouped in
clusters of \ such collections, which reasonably can be considered
as being weakly bounded \ among them. This structure suggests that
the pressureless condition of the gas of galaxies is a reasonable
assumption.

\section{Field equations}

\ Given the isotropic and stationary character of the solution which is
searched, let us propose the structure of the metric in the standard form
\begin{align}
ds^{2} &  =\mathit{v}(\rho){dx^{o}}^{2}-u(\rho)^{-1}d\rho^{2}-\rho^{2}%
(sin^{2}\theta d\varphi^{2}+d\theta^{2}),\nonumber\\
x^{o} &  =ct\text{, \ \ \ }x^{1}=\rho,\\
x^{2} &  =\varphi,\text{ \ }x^{3}\equiv\theta,
\end{align}
from which the components of the Einstein tensor $G_{\mu\nu}$ can be
computed. Since the metric tensor is diagonal and only depending on
$\rho$, the only non vanishing components of \ $G_{\mu\nu}$ result
in
\begin{align*}
{G_{0}^{0}} &  ={\frac{u^{\prime}}{\rho}}-{\frac{1-u}{\rho^{2}}},\\
G_{1}^{1} &  ={\frac{u}{v}}{\frac{v^{\prime}}{\rho}}-{\frac{{1-u}}{\rho^{2}}%
},\\
{G_{2}^{2}}={G_{3}^{3}} &  ={\frac{u}{2v}}{v^{\prime\prime}%
}+{\frac{uv^{\prime}}{4v}(\frac{u^{\prime}}{u}-\frac{v^{\prime}}{v})}\\
&  +{\frac{u}{2\rho}(\frac{u^{\prime}}{u}+\frac{v^{\prime}}{v})}.
\end{align*}
The components $G_{2}^{2}$ and $G_{3}^{3}$ generate second order equations in
the temporal component of the metric, which explicitly do not play an
important role thanks to the Bianchi \ identities \cite{Weinberg}:
\begin{equation}
{G}_{\mu;{\nu}}^{\nu}=0.\label{bianchi1}%
\end{equation}
which will be employed below. \ Assumed the satisfaction of the Einstein
equations the $G_{\mu}^{\nu}$ tensor can be substituted by the energy momentum
tensor $T_{\mu}^{\nu}$. Equation (\ref{bianchi1}) is interpreted a set of
dynamical equations for the parameters, that is $e$, $p$ and $\phi$.

\subsection{Matter and Dilaton Dark Energy}

In this section let us sketch the way followed for obtaining two of the
necessary equations needed to show the existence of the mentioned static model
for the Universe:  the Bianchi relations (\ref{bianchi1}) and the static
equation for the scalar field coupled to matter. .

Let us write the action for the scalar field-matter in the given \ space time
in the form
\begin{equation}
{S}_{mat-\phi}=\int L\sqrt{-g}d^{4}x,
\end{equation}
where $g$ is the determinant of the metric tensor,  and consider that the
Lagrangian density takes the form:
\begin{equation}
{L}={\frac{1}{2}}(g^{\alpha\beta}{\phi}_{,\alpha}{\phi}_{,\beta}+m^{2}{\phi
}^{2})+j\phi+L_{e,p}.\label{denslag}%
\end{equation}
The first and the third terms of the right member of (\ref{denslag})
\ are the Lagrangian densities of the  KG scalar field and the
dust-like matter respectively; while the second term is an
interaction term between both which is added. The strength of the
interaction will be represented by the constant source $j$.

Based in the argue given in the last paragraph of section 1, we will consider
for the matter the perfect fluid expression \cite{Weinberg}:
\begin{equation}
(T_{e,p})_{\mu}^{\nu}=p\,\delta_{\mu}^{\nu}+u^{\nu}u_{\mu}(p+e),
\end{equation}
where  $p$ is the  pressure  of the matter. Note that  \ we will
assume pressure-less matter $p=0$. However, for bookkeeping
purposes, we will get the expression for a general pressure $p$ up
to the end when the limit $p=0$ will be fixed.

As usual $u^{\nu}$ denote the contra-variant components of the 4-velocity of
the fluid in the system of \ reference under consideration. \ In addition
since we seach for static configurations the 4-velocity takes the simple form
$u^{\nu}=\delta_{0}^{\nu}$.

From the Lagrangian $L$ en (\ref{denslag}) and the above remarks \ the energy
momentum tensor of the scalar field coupled with the matter gets the form%
\begin{align}
T^{\nu}_{\mu}  &  =-\frac{\delta^{\nu}_{\mu}}{2}(g^{\alpha\beta}{\phi
}_{,\alpha}{\phi}_{,\beta}+m^{2}{\phi}^{2}+2j \,\phi)\nonumber\\
&  +g^{\alpha\nu}{\phi}_{,\alpha}{\phi}_{,\mu}+p\delta^{\nu}_{\mu}+\delta
^{\nu}_{0}\delta_{\mu}^{0}(p+e).   \label{tmunu}%
\end{align}

From equation (\ref{tmunu}), the Bianchi relation for $\mu=1$ en
(\ref{bianchi1}) \ transforms in
\[
-\phi j^{\prime}+p^{\prime}+\frac{v^{\prime}}{2v}(p+e)=0.
\]
In our case this is the only  of the four Bianchi relations which is different
from zero.

The dynamical equation for the scalar field which determines the extremum of
the action $S_{mat-\phi}$, takes the form
\begin{align}
\frac{\delta S_{mat-\phi}}{\delta\phi} &  \equiv\frac{D}{dx^{\mu}}%
\frac{\partial L}{\partial\phi_{,\mu}}-\frac{\partial L}{\partial\phi
}\nonumber\\
&  \equiv\frac{1}{\sqrt{-g}}\frac{\partial}{\partial x^{\mu}}(\sqrt{-g}%
g^{\mu\nu}\phi_{,\nu})-m^{2}\phi-j\nonumber\\
&  =0,
\end{align}
which after introducing the components of the metric tensor
simplifies to
\begin{equation}
u\phi^{\prime\prime}+u\phi^{\prime}(\frac{1}{2}\frac{v^{\prime}}{v}+\frac
{1}{2}\frac{u^{\prime}}{u}+\frac{2}{\rho})-m^{2}\phi
-j=0.\nonumber\label{escalar}%
\end{equation}
Note that if $u=v=1$, that is, in the Minkowski space,  relation
(\ref{escalar}) reduces to the static KG equation for scalar field interacting
with an external source $j$. \ It might be helpful to notice that natural
units
\[
\lbrack e]=[p]=cm^{-4},[m]=cm^{-1},[\phi]=cm^{-1}%
\]
are employed.

\subsection{Einstein equations}

The extremum of the action ${S}_{mat-\phi}$with respect to the
metric leads to the Einstein equations in the absence of a
Cosmological Constant
\begin{equation}
G_{\mu}^{\nu}=G\hspace{0.1mm}T_{\mu}^{\nu},\label{einecua0}%
\end{equation}
where in natural units $G=8\pi\times l_{p}^{2}$ \ and $l_{p}=1.61\times
10^{-33}cm$ is the Planck length.

From relation  (\ref{tmunu}), the Einstein equations
(\ref{einecua0}) \ take the form%
\begin{align}
{\frac{u^{\prime}}{\rho}}-{\frac{1-u}{\rho^{2}}}  &  =-G[\frac{1}{2}%
(u\phi_{,\rho}^{2}+m^{2}\phi^{2}+2j\phi)+e],\label{ee1}\\
{\frac{u}{v}}{\frac{v^{\prime}}{\rho}}-{\frac{{1-u}}{\rho^{2}}}  &
=G[\frac{1}{2}(u\phi_{,\rho}^{2}-m^{2}\phi^{2}-2j\phi)+p\,] . \label{ee2}%
\end{align}

As it was mentioned above, the third Einstein equation is not needed
for determining the solution, because its satisfaction is implied by
the other equations. This expression only imposes the continuity of
the of the derivative of $v$ \ with respect to the radial variable
since it is a second order differential equation.

We will assume that $j$  which gives the form of the interaction
term between the dark energy and matter \ is of the form:
\[
j=ge^{\frac{1}{2}},
\]
where $g$ is a coupling constant for the interaction matter-scalar field. In
the natural system of units $[g]=cm^{-1}$. \

\subsection{Working equations}

With the aim of \ working with dimensionless forms of the equations
(\ref{ee1}) and (\ref{ee2}), let us define the new variables and parameters
\begin{align*}
r\equiv m\rho,\text{ \ } &  \text{ \ }\Phi\equiv\sqrt{8\pi}l_{p}\phi,\\
J\equiv\frac{\sqrt{8\pi}l_{p}}{m^{2}}j, &  \text{ \ \ }\epsilon\equiv
\frac{8\pi l_{p}^{2}}{m^{2}}e,\text{ \ \ \ }\gamma\equiv\frac{g}{m}.
\end{align*}
Let us now fix the mass of the Dilaton field \ to the value
estimated in Ref. \ \cite{papa1} for assuring the observed strength
of the  Hubble effect in the regions near the origin. \
Interestingly, this value resulted in the very small quantity, $\ \
m=4\times10^{-29}cm^{-1}$ . This mass is compatible with the zero
mass Dilaton \ in the lowest approximation. Also the mass is of the
order of the inverse of the estimated radius of the Universe, like
it was observed in Ref. \cite{papa1}.

Therefore,  the set of working equations will be
\begin{align}
{\frac{u_{,r}}{r}}-{\frac{1-u}{r^{2}}} &  =-\frac{1}{2}(u{\Phi_{,r}}^{2}%
+\Phi^{2})-J\Phi-\epsilon,\label{eecuaadim1}\\
\frac{u}{v}\frac{v_{,r}}{r}-{\frac{1-u}{r^{2}}} &  =-\frac{1}{2}(-u{\Phi_{,r}%
}^{2}+\Phi^{2}+2J\Phi),\label{eecuaadim2}\\
\epsilon\frac{v_{,r}}{2v}-\Phi J_{,r} &  =0,\label{eecuaadim3}\\
u\Phi_{,rr}-\Phi-J &  =-u\Phi_{,r}(\frac{1}{2}\frac{v_{,r}}{v}+\frac{1}%
{2}\frac{u_{,r}}{u}+\frac{2}{r}).\label{eecuaadim4}%
\end{align}

\subsection{The solutions near\ the center of symmetry}

We will search for  smooth solutions around the origin. Thus, the
continuity of the derivatives $v$ and $\phi$, in all places
including the origin, will be required. \ Thus considering the
equations in a neighborhood of the origin \ the asymptotic field
values can be got in the form
\begin{align*}
u &  =1+u_{1}r^{2}...,\\
v &  =1+v_{1}r^{2}...,\\
\Phi &  =\Phi_{0}+\Phi_{1}r^{2}...,\\
\epsilon &  =\epsilon_{0}+\epsilon_{1}r^{2}... ,
\end{align*}
where $u_{1}$, $v_{1}$, $\Phi_{1}$, $\epsilon_{1}$ are given by the relations
\begin{align}
u_{1} &  =-\frac{1}{3}(\frac{\Phi_{0}^{2}}{2}+J_{0}\Phi_{0}+\epsilon_{0}),\\
v_{1} &  =-\frac{1}{3}(\frac{\Phi_{0}^{2}}{2}+J_{0}\Phi_{0}-\frac{\epsilon
_{0}}{2}),\\
\Phi_{1} &  =-\frac{1}{6}(\Phi_{0}+J_{0}),\\
\epsilon1 &  =-\frac{\epsilon_{0}^{\frac{3}{2}}}{3\gamma\Phi_{0}}(\frac
{\Phi_{0}^{2}}{2}+J_{0}\Phi_{0}-\frac{\epsilon_{0}}{2}),\\
J_{0} &  =\gamma\epsilon_{0}^{\frac{1}{2}}.\nonumber
\end{align}

Note that the spacial dependence of the metric has an \ exact homogeneous
structure near the center of symmetry. \ The quantities $\Phi_{0}$,
$\epsilon_{0}$ and the dimensionless coupling constant $\gamma$ remains as
free parameters.  Future extensions of this work, are considered to optimize
the parameters aiming to compare the predictions of the model with redshift
vs.estelar magnitude in the  supernovae obervations.  In what follows we only
will illustrate the general behavior of the solutions for some physically
motivated values \ of the parameters. . \

\subsection{Solutions}

\begin{figure}[ptb]
\par
\begin{center}
\epsfig{figure=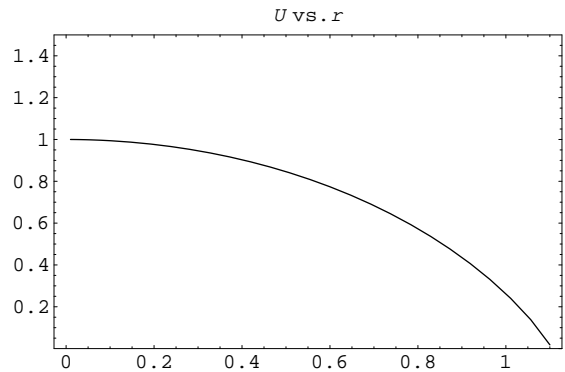,width=8cm} \vspace{-0cm}
\end{center}
\caption{The radial contraviant component of the metric $g^{11}\equiv u(r)$
behaves basically as the as the as in the deSitter Universe having the size
$R\equiv$0.25$\times10^{29}$cm.}%
\label{grafico1}%
\end{figure}
\begin{figure}[ptb]
\par
\begin{center}
\epsfig{figure=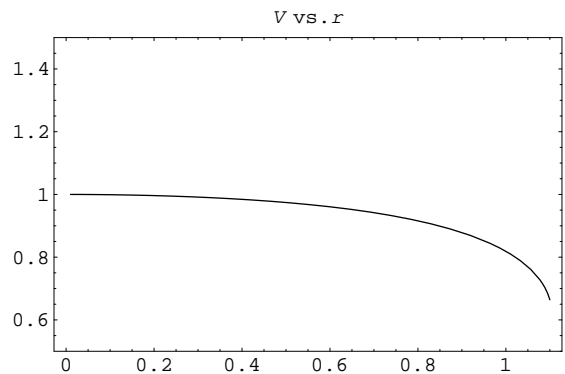,width=8cm} \vspace{-0cm}
\end{center}
\caption{Temporal component of the metric $g_{00}\equiv$v(r). Its decreasing
beahvour show the redshift of the arriving to the central regions form the far
regions. The radius of the singularity at the far away region is $R$ $\equiv
$0.25$\times10^{29}$cm.}%
\label{grafico2}%
\end{figure}

\begin{figure}[ptb]
\par
\begin{center}
\epsfig{figure=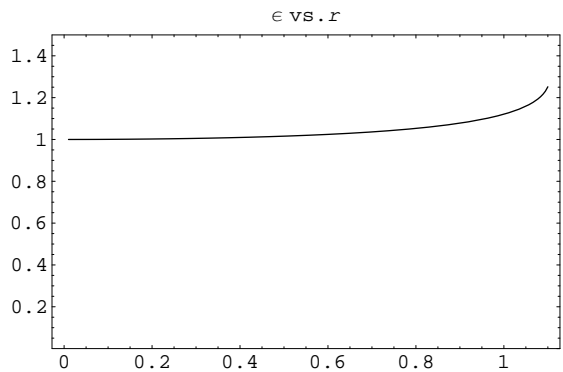,width=8cm} \vspace{-0cm}
\end{center}
\caption{The matter distribution $e(r)$ is slowlly varying witht he radial
distance. The coupling between the scalar field and the matter $J\Phi$ is
central in allowing the existence of the static solution, in which also the
matter to Dark energy content ratio is also slowly varying. The radial
singularity defining the end of the space time is at $R=0.25\times10^{29}$cm.}%
\label{grafico3}%
\end{figure}

\begin{figure}[ptb]
\epsfig{figure=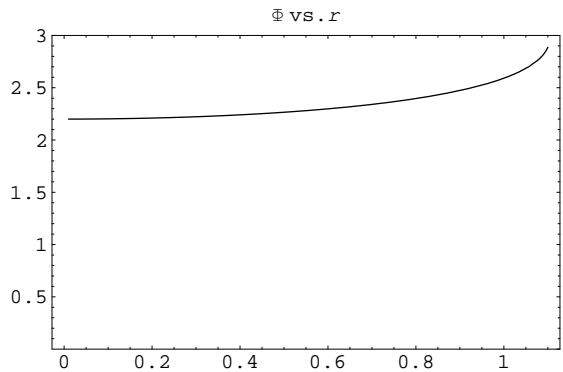,width=8cm} \vspace{-0cm}
\caption{The scalar field slowly varies with the radial component
and behaves very closely with the matter density $e(r)$; The radial
singularity defining the end of the space time is at
$R=0.25\times10^{29}$cm. There is no static metric
with Dilaton and matter in coexistence without interaction. }%
\label{grafico4}%
\end{figure}

Let us consider the numerical solutions of the equations
(\ref{eecuaadim1})-(\ref{eecuaadim4}), selecting the parameter
values $\gamma=-0.75$, $\Phi_{0}=2.2$ and $\epsilon_{0}=1.$ These
specific values correspond to \ a coupling constant
$g=2.9\times10^{-29}cm^{-1}$, a value of the scalar field at the
origin $\phi_{0}=2.7\times10^{32}cm^{-1}$ (that is at the Planck \
scale) \ and at a matter energy density of $e=2.3\times
10^{7}cm^{-4}.$ The numerical solutions of the \ equations
(\ref{eecuaadim1})-(\ref{eecuaadim4}) are illustrated in the figures
(\ref{grafico1})-(\ref{grafico4}).

These parameters  were a priori selected with the aim of fixing the
estimated value of $0.7/0.3$ for the ratio of the Dark Energy to the
matter energy content  in the Universe \cite{pront}and the  the
approximate value of the \ Hubble effect.

From Fig..(\ref{grafico1})  the global similarity between the space -time
being studied and the de Sitter static solution can be observed. Moreover, due
to the chosen value of the Dilaton mass suggested in (\cite{papa1}), \ the
size  of the Universe (defined as the radial distance at which the singularity
of the structure appears) \ is of  the order of the estimated \ value
$10^{29}cm$. In Fig.(\ref{grafico2}) the dependence form of the temporal metric
is shown, it evidences that the observer near the origin measure a redshift
\ which was fixed to have a value being near to the observed now.

Figures (\ref{grafico3}) and (\ref{grafico4}) illustrate the distribution of
\ energy and scalar field respectively \ Note the similarity between both
magnitudes. That is,  the existence of the coupling not only allows the
existence of the static solution, but in addition it also produces  a
configuration in which the proportion of matter and dark energy is more or
less approximately constant over large regions of the space time. \ \

\begin{acknowledgments}
The invitation and kind hospitality of the High Energy Section of
the Abdus Salam International Center for Theoretical Physics
(ASICTP) and its Head S. Randjbar-Daemi, allowing for a very
helpful visit to the Center, is deeply acknowledged. I express
also my gratitude by the support to the work received from the
Office of External Activities of ICTP (OEA),  through the Network
on \textit{Quantum Mechanics, Particles and Fields}(Net-35).  The
useful remarks during the stay at the AS ICTP received from P. Creminelli 
are also very  much appreciated.
\end{acknowledgments}


\end{document}